\documentclass{aa}

\usepackage{graphicx}
\usepackage{natbib}
\bibpunct{(}{)}{;}{a}{}{,}

\title{
The evolution of the ultraviolet and infrared luminosity densities 
in the universe at $0<z<1$
}

\author{
  T.~T.~Takeuchi\thanks{
  Postdoctoral Fellow of the Japan Society for the Promotion of Science (JSPS)
  for Research Abroad.
}
  \and 
  V.~Buat
  \and
  D.~Burgarella
}

\institute{Laboratoire d'Astrophysique de Marseille, 
  Traverse du Siphon BP8, 13376 Marseille Cedex 12, France\\
  \email{[tsutomu.takeuchi,veronique.buat,denis.burgarella]@oamp.fr}
}

\offprints{T. T. Takeuchi\\
\email{tsutomu.takeuchi@oamp.fr}.
}

\date{Received/Accepted}

\titlerunning{Luminosity density evolution in UV and IR}
\authorrunning{T.\ T.\ Takeuchi et al.}

\begin{document}

\abstract{
The ratio between far-ultraviolet (FUV) and infrared (IR) luminosity 
densities from $z=0$ to $z=1$ is discussed by using the luminosity functions
(LFs) of both wavelengths.
The FUV LF ($z=0\mbox{--}1$) based on {\sl GALEX} has been reported by
\citet{arnouts05}, whilst for the IR LF, we used the {\sl IRAS} 
PSC$z$ 60-$\mu$m LF for the local universe and the {\sl Spitzer} 
15-$\mu$m LF at higher-$z$ as used by \citet{lefloch05}.
Both luminosity densities show a significant evolutionary trend, but the IR 
evolves much faster than the FUV.
Consequently, the ratio $\rho_{\rm dust}/\rho_{\rm FUV}$ increases toward
higher-$z$, from $\sim 4$ (local) to $\sim 15$ ($z\simeq 1$).
It is also shown that more than $70\;\%$ of the star formation activity in 
the universe is obscured by dust at $0.5 \la z \la 1.2$.
}

\maketitle

\section{Introduction}\label{sec:introduction}

Dust attenuation is one of the most fundamental obstacles when we study the 
star formation activity of galaxies.
Since ultraviolet (UV) radiation is emitted by young stars, it is in principle
directly related to the recent star formation rate (SFR).
However, the use of UV light to trace the SFR is strongly hampered by the 
presence of dust, which absorbs and scatters the UV photons and finally 
re-emits the energy in the IR (mainly far-IR: FIR).
Therefore, the UV and IR emissions play complementary roles in estimating
the recent SFR of galaxies.

In particular, the effect of dust has given rise to a long lasting debate 
on the cosmic star formation history \citep[e.g.,][]{hopkins04}.
The most direct way to address this issue is to compare the observed 
cosmic luminosity densities in UV and IR.
In such studies, the luminosity function (LF) of galaxies is the starting 
point \citep[e.g.,][]{takeuchi00}.
In this work, we investigate the evolution of FUV and IR LFs.

A highly reliable LF in the FUV has recently been published based on new 
UV data obtained by 
{\sl GALEX}\footnote{URL: {\tt http://http://www.galex.caltech.edu/}.} 
\citep[e.g.,][]{arnouts05}.
In the IR, \citet{takeuchi03b} constructed a 60~$\mu$m LF from {\sl IRAS} 
PSC$z$ \citep{saunders00}.
Hereafter, we indicate IR emission integrated over $8\mbox{--}1000\;\mu$m 
by a subscript `dust'.
For higher-$z$, first results from {\sl Spitzer}\footnote{URL: 
{\tt http://www.spitzer.caltech.edu/}.}
have been recently reported \citep{lefloch05,perez_gonzalez05}.
Making use of these LFs, we calculate the luminosity density at both 
wavelengths to examine the SFR history in the universe at $0<z<1$.

Throughout this manuscript, we adopt a flat lambda-dominated cosmology with 
$(h,\Omega_0,\lambda_0)=(0.7,0.3,0.7)$, where $h\equiv H_0/100 
[\mbox{km\,s}^{-1}\mbox{Mpc}^{-1}]$, $\Omega_0$ is the density parameter, 
and $\lambda_0$ is the normalized cosmological constant.

\section{Luminosity functions}\label{sec:lf}

We define the luminosity function as a number density of galaxies whose
luminosity lies between a logarithmic interval
$[\log L, \log L + d\log L]$:\footnote{We denote $\log x \equiv \log_{10} x$
and $\ln x \equiv \log_e x$.} $\phi (L) \equiv dn/d \log L$.
Here we define the luminosity at a certain wavelength band by 
$L\equiv\nu L_\nu$.

\subsection{The FUV luminosity function}

\begin{table}
\centering
 \caption{Schechter parameters for FUV luminosity function 
\citep[converted from][]{arnouts05}.}
\label{tab:param_uv}
  \begin{tabular}{lccc}
   \hline
    Redshift & $\alpha$ & $L_*(\mbox{FUV})$     & $\phi_*$          \\
             &          & $h^{-2}\; [L_\odot] $ & $h^3\;[\mbox{Mpc}^{-3}]$ \\
   \hline
    0        & 1.21     & $1.81 \times 10^{9} $ & $1.35 \times 10^{-2} $   \\
    0.2--0.4 & 1.19     & $2.43 \times 10^{9} $ & $1.95 \times 10^{-2} $   \\
    0.4--0.6 & 1.55     & $6.75 \times 10^{9} $ & $5.38 \times 10^{-3} $   \\
    0.6--0.8 & 1.60     & $9.32 \times 10^{9} $ & $5.25 \times 10^{-3} $   \\
    0.8--1.2 & 1.63     & $1.19 \times 10^{10}$ & $3.63 \times 10^{-3} $   \\
   \hline
 \end{tabular}
\end{table}
\citet{wyder05} estimated the local LF of galaxies at FUV ($1530$~\AA) from
{\sl GALEX} data in overlapped regions with 2dFGRS \citep[][]{colless01}.
The local FUV LF is well described by a Schechter function \citep{schechter76}
\begin{eqnarray}\label{eq:schechter}
  \phi (L) = (\ln 10)\; \phi_* \left( \frac{L}{L_*} \right)^{1-\alpha}
  \exp \left[-\left(\frac{L}{L_*}\right)\right]\;,
\end{eqnarray}
At $z=0$, $(\alpha, L_*, \phi_*) = (1.21, 1.81\times 10^9h^{-2}\;L\odot, 
1.35 \times 10^{-2}h^3\;\mbox{Mpc}^{-3})$.

\citet{arnouts05} presented the evolution of the {\sl GALEX} FUV LF 
using the VIRMOS VLT Deep Survey \citep[VVDS: see,][]{lefevre03}.
They found that the FUV LFs at higher $z$ are also well fitted by the 
Schechter function.
\citet{arnouts05} directly measured the parameters 
($\alpha, L_*, \phi_*$) at each redshift bin.
They reported that the $\alpha$ and $L_*$ monotonically increase with $z$,
while the $\phi_*$ decreases with $z$.
We adopt the converted values of the parameters in 
Equation~(\ref{eq:schechter}) (Table~\ref{tab:param_uv}).

\subsection{The IR luminosity function}

Contrary to the FUV LF, the local IR LF is not well-fitted by 
a Schechter function, although it can be expressed as a function given by
\citet{saunders90} which is defined as
\begin{eqnarray}\label{eq:saunders}
  \phi (L) = \phi_* \left( \frac{L}{L_*} \right)^{1-\alpha}
    \exp \left\{ -\frac{1}{2\sigma^2} 
    \left[ \log \left(1+\frac{L}{L_*}\right)\right]^2\right\}\;.
\end{eqnarray}

We use the parameters for the local $60\;\mu$m LF, given by 
\citet{takeuchi03b}.
We converted the $60\;\mu$m luminosity to that of the total dust
emission by adopting the average 60$\mu$m-to-dust flux ratio of $2.5$ 
estimated from the PSC$z$ sample of \citet{takeuchi05a}.
This luminosity-independent conversion can be justified because of 
the tight linear correlation of the two quantities
(correlation coefficient $r=0.991$).

Since most of the galaxies in {\sl IRAS} PSC$z$ are local ($z<0.1$),
we need a deep survey result to evaluate the evolution of the IR LF.
Recently, very important results from {\sl Spitzer} MIPS 24-$\mu$m 
observations have been reported on the mid-IR (MIR: 12 or $15\;\mu$m) LF 
\citep{lefloch05,perez_gonzalez05}.
These authors reported a very strong evolution trend for the MIR LF.
Hereafter we adopt \citet{lefloch05} because their LFs are given in a
form identical to the one adopted by us.
However, we remark that \citet{perez_gonzalez05}, although adopting a 
different form to our study and that of \citet{lefloch05}, nonetheless
also reached consistent conclusions on the amount of the evolution.

\citet{lefloch05} first estimated nonparametric 15-$\mu$m LFs for each 
redshift bin at $0 < z < 1$ from MIPS 24-$\mu$m data by the $1/V_{\rm max}$ 
method.
Then, using some spectral energy distribution (SED) templates, they converted
their 15-$\mu$m luminosity to the total dust luminosity.
Based on these nonparametric LFs at $0 < z < 1$, they estimated
the evolution rate of the IR LF, by adopting the form 
\begin{eqnarray}\label{eq:evol_formula}
  \phi(L,z) = g(z)\phi_0\left[\frac{L}{f(z)}\right] \;,
\end{eqnarray}
where $\phi_0(L)$ is the local functional form of the LF.
They assumed a power-law form for the evolution functions $f(z)$ and $g(z)$ as
\begin{eqnarray}
  f(z)=(1+z)^{Q} \; , \quad 
  g(z)=(1+z)^{P} \; .\label{eq:evol_qp}
\end{eqnarray}
Through a $\chi^2$ minimization between the nonparametric LFs and 
Equation~(\ref{eq:evol_formula}), they obtained the parameter values as
$Q=3.2^{+0.7}_{-0.2}$ and $P=0.7^{+0.2}_{-0.6}$.
This means that $L_* \propto (1+z)^{3.2}$ and 
$\phi_* \propto (1+z)^{0.7}$ in Equation~(\ref{eq:saunders}), 
whilst $\alpha$ remains constant.

Here we mention the uncertainty in the conversion of luminosities.
\citet{lefloch05} converted the local 60-$\mu$m LF of \citet{takeuchi03b} to 
the dust LF by using their model SED templates.
Since their conversion procedure is different from ours, we examined their
consistency.
The difference between their conversion and ours does not exceed 5-\%,
thus we judge the difference to be negligible for the subsequent analysis 
at $60\;\mu$m for $z=0$.

At higher-$z$, they convert $L_{15}$ to $L_{\rm dust}$ by SED templates.
We also consider the potential systematic uncertainty introduced by this
procedure.
In particular, the evolution of the population causes an increase of the 
fraction of IR luminous galaxies (LIRGs, ULIRGs), which may have different 
SEDs to less active galaxies.
This can lead a systematic change of the corresponding template SED with $z$.
{}To evaluate this uncertainty, \citet{lefloch05} made a comparison between 
several IR SED templates 
\citep{dale01,chary01,dale02,chanial03,lagache03,lagache04}.
{}From their Fig.~8, they estimated the typical uncertainty to be 
$\sim 0.2\;\mbox{dex}$.
We further extended their test using some other SED libraries
\citep{efstathiou00,efstathiou03,takeuchi01a,takeuchi01b}, and confirm their
claim.
The good linearity between MIR ({\sl IRAS} 12 and $25 \;\mu$m) and 
dust luminosities at a very wide range of luminosity 
($10^{6}\;L_\odot \mbox{--} 10^{13}\;L_\odot$)
\citep{takeuchi05a}, even for galaxies with extreme SEDs 
\citep[see, e.g.,][]{takeuchi03a,takeuchi05b}, also ensures the robustness 
of the estimation.
In summary, the uncertainty of the dust luminosity estimation is a factor of
three.
We should keep this uncertainty in mind for the following.

\section{Result and discussion}

\subsection{Evolution of the luminosity contribution in the FUV and IR}

\begin{figure}
\centering\includegraphics[width=0.45\textwidth]{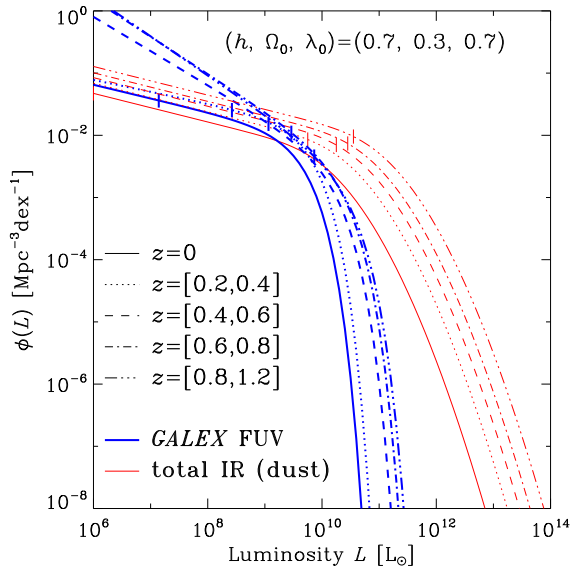}
\caption{The evolution of the luminosity function (LF) of galaxies 
in the far-ultraviolet (FUV: 1530~\AA) obtained by {\sl GALEX} 
and infrared (dust: $8\mbox{--}1000\;\mu$m) obtained from {\sl IRAS} PSC$z$ 
($z=0$) and {\sl Spitzer} at higher-$z$.
Thick lines show the LFs in the FUV, and thin lines depict those of dust.
Vertical tick marks on the LF indicate the lowest luminosity limits 
above which the observed data exist.
}\label{fig:compare_lf}
\end{figure}

\begin{figure}
\centering\includegraphics[width=0.45\textwidth]{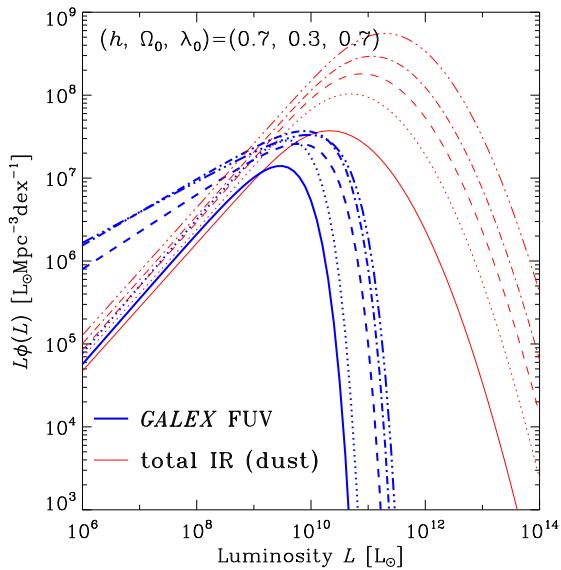}
\caption{The evolution of the contribution of galaxies to the luminosity 
density in the FUV and IR.
The meaning of the different lines are the same as that of 
Figure~\ref{fig:compare_lf}.
}\label{fig:compare_lphi}
\end{figure}

The evolution of the FUV and dust LFs are shown in 
Figure~\ref{fig:compare_lf}.
First, it is worth mentioning the (well-known) difference of the local 
LF shape of FUV and dust: for the dust LF, bright galaxies 
($L \ga 10^{10}\;L_\odot$) are much more numerous than those in the FUV.
This leads to the difference in the main population contributing to the total
emitted energy.
The product $L\phi(L)$ represents the energy contribution of galaxies with
luminosity $L$ (see Figure~\ref{fig:compare_lphi}).
In the FUV, the main contributor is $L_*$ galaxies, with fainter galaxies
emitting a non-negligible fraction of energy at $z>0.5$.
In contrast, the effect of the evolution appears in the bright end
for the dust LF.
The contribution from the most luminous galaxies increases with redshift.

\subsection{Evolution of the FUV and dust luminosity densities and 
the mean dust attenuation}

\begin{table*}
\centering
 \caption{Evolution of the FUV and dust luminosity densities, the mean dust 
attenuation, and the converted cosmic SFR densities.}
\label{tab:lum_density}
  \begin{tabular}{lccccccccc}
   \hline
    Redshift 
    & $\rho_{\rm FUV}$
    & $\rho_{\rm dust}$ 
    & $\rho_{\rm dust}/\rho_{\rm FUV}$
    & $A(\mbox{FUV})$
    & $\rho_{\rm SFR}(\mbox{FUV})$ 
    & $\rho_{\rm SFR}(\mbox{dust})$
    & Hidden SFR \\

    & $h\;[L_\odot \mbox{Mpc}^{-3}]$ 
    & $h\;[L_\odot \mbox{Mpc}^{-3}]$ 
    & 
    & [mag]
    & $h\;[M_\odot\mbox{yr}^{-1}\mbox{Mpc}^{-3}]$ 
    & $h\;[M_\odot\mbox{yr}^{-1}\mbox{Mpc}^{-3}]$
    & fraction \\
   \hline
    0        & $2.71\times 10^{7}$ &  $1.02^{+0.59}_{-0.37}\times 10^{8}$
    & $3.75^{+2.19}_{-1.38}$ & $1.29^{+0.32}_{-0.30}$ & $8.37 \times 10^{-3}$ 
    & $ 1.08^{+0.63}_{-0.40}\times 10^{-2}$ & $0.56^{+0.11}_{-0.11}$  \\
    0.2--0.4 & $5.44\times 10^{7}$ &  $2.83^{+1.65}_{-1.82}\times 10^{8}$
    & $5.19^{+3.04}_{-1.92}$ & $1.52^{+0.34}_{-0.32}$ & $1.68 \times 10^{-2}$ 
    & $ 3.02^{+1.76}_{-1.11}\times 10^{-2}$ & $0.64^{+0.10}_{-0.11}$  \\
    0.4--0.6 & $6.96\times 10^{7}$ &  $4.94^{+2.89}_{-1.82}\times 10^{8}$
    & $7.09^{+4.14}_{-2.62}$ & $1.75^{+0.35}_{-0.33}$ & $2.15 \times 10^{-2}$
    & $ 5.27^{+3.08}_{-1.94}\times 10^{-2}$ & $0.71^{+0.09}_{-0.10}$  \\
    0.6--0.8 & $1.06\times 10^{8}$ &  $8.05^{+4.71}_{-2.97}\times 10^{8}$
    & $7.58^{+4.43}_{-2.80}$ & $1.80^{+0.36}_{-0.34}$ & $3.28 \times 10^{-2}$ 
    & $ 8.59^{+5.02}_{-3.17}\times 10^{-1}$ & $0.72^{+0.08}_{-0.10}$  \\
    0.8--1.2 & $1.00\times 10^{8}$ &  $1.52^{+0.89}_{-0.56}\times 10^{9}$
    & $15.1^{+8.86}_{-5.59}$ & $2.34^{+0.39}_{-0.37}$ & $3.10 \times 10^{-2}$
    & $ 1.62^{+0.95}_{-0.60}\times 10^{-1}$ & $0.84^{+0.05}_{-0.07}$  \\
   \hline
 \end{tabular}
\end{table*}

For FUV, we integrate $L\phi(L)$ over $L_{\rm FUV} = 10^6\;L_\odot
\mbox{--}10^{15}\;L_\odot$ at each $z$ to obtain the evolution
of the FUV luminosity density $\rho_{\rm FUV}$, while
for the dust, we first construct a LF at a given $z$ according to 
Equations~(\ref{eq:evol_formula}), and (\ref{eq:evol_qp})
with the estimated value of \citet{lefloch05}, and integrate $L\phi(L)$ over 
the same range as that of FUV galaxies.
The densities depend very little on the integration luminosity range:
even if we change the lower bound to $1\;L_\odot$, the integrated value
only increases by 0.4~\% for the dust luminosity and by 3~\% in the FUV 
luminosity.

The luminosity densities are summarized in Table~\ref{tab:lum_density}.
Both luminosity densities show a significant evolutionary trend, but the dust 
luminosity density evolves much faster than that of the FUV.
In Table~\ref{tab:lum_density}, we only list the systematic uncertainty
potentially introduced by the choice of SED templates.
For statistical errors, see \citet{schiminovich05} for the FUV and 
\citet{lefloch05} for $15\;\mu$m.
Consequently, the ratio $\rho_{\rm dust}/\rho_{\rm FUV}$ increases toward
higher-$z$, from $\sim 4$ (local) to $\sim 15$ ($z\simeq 1$), 
i.e., the dust luminosity dominates the universe at $z\sim 1$.

The FUV-to-dust luminosity density ratio can be interpreted as the global
mean dust attenuation in the universe.
\citet{buat05} provided a formula which relates the dust to FUV flux 
ratio to the dust attenuation in the FUV, $A(\mbox{FUV})$~[mag], as
\begin{eqnarray}\label{eq:attenuation}
  A(\mbox{FUV})=-0.0333y^3+0.3522y^2+1.1960y+0.4967 \;,
\end{eqnarray}
where $y\equiv \log \left( F_{\rm dust}/F_{\rm FUV}\right)$
and $F=\nu S_\nu$ ($S_\nu$: flux density) .
The mean attenuation obtained by Equation~(\ref{eq:attenuation}) is also 
tabulated in Table~\ref{tab:lum_density}.

\subsection{Fraction of obscured star formation}

\begin{figure}
\centering\includegraphics[width=0.45\textwidth]{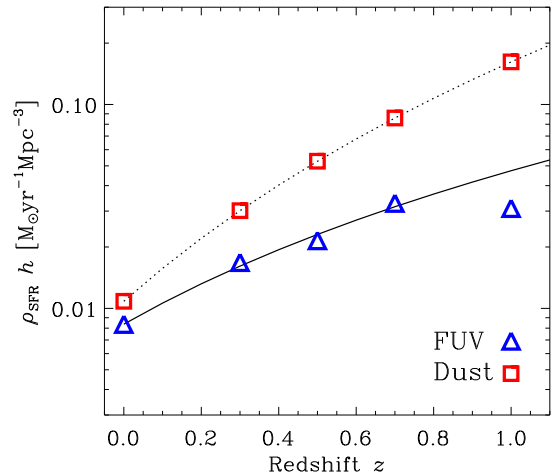}
\caption{The evolution of the star formation densities derived from FUV and
dust.
Typical statistical error is about a factor of two, and systematic 
uncertainty related to IR SED template is about $\sim 0.2\;\mbox{dex}$.
}\label{fig:sfr}
\end{figure}

We interpret the data in terms of SFR.
Assuming a constant SFR over $10^8$~yr, and Salpeter initial mass function 
(IMF) \citep[][mass range: $0.1\mbox{--}100\;M_\odot$]{salpeter55}, 
Starburst99 \citep{leitherer99} gives the relation between the SFR and 
$L(\mbox{FUV})\equiv \nu L_\nu$ at FUV (1530~\AA),
\begin{eqnarray}\label{eq:conv_SFR_uv}
  \log L(\mbox{FUV}) = 9.51 + \log \mbox{SFR} \;.
\end{eqnarray}
For the IR, to transform the dust emission to the SFR, we assume
that all the stellar light is absorbed by dust.
Then, we obtain the following formula under the same assumption for both 
the SFR history and the IMF as those of the FUV, 
\begin{eqnarray}\label{eq:conv_SFR_ir}
  \log L(\mbox{dust}) = 9.75 + \log \mbox{SFR} \;.
\end{eqnarray}
However, a significant fraction of the dust emission is due to the heating of 
grains by old stars which is not directly related to the recent SFR.
\citet{hirashita03} found that about 40~\% of the dust heating in the nearby
galaxies comes from stars older than $10^8$~yr.
Adopting this correction, we obtained the evolution of the star formation 
rate densities from FUV and dust ($\rho_{\rm SFR}({\rm FUV})$ and 
$\rho_{\rm SFR}({\rm dust})$) which are presented in 
Figure~\ref{fig:sfr} and Table~\ref{tab:lum_density}.

The evolution of $\rho_{\rm dust}$ and $\rho_{\rm FUV}$ (therefore 
$\rho_{\rm SFR}(\mbox{dust})$ and $\rho_{\rm SFR} (\mbox{FUV})$)
is described by the form of $(1+z)^R$.
The power-law index $R$ of $\rho_{\rm dust}$ is estimated to be $3.9 \pm 0.4$ 
by \citet{lefloch05}, and for 
$\rho_{\rm FUV}$, \citet{schiminovich05} give an index of $2.5 \pm 0.7$
(dotted and solid curves in Figure~\ref{fig:sfr}).
The evolution of $\rho_{\rm dust}$ is slightly stronger than 
suggested by \citet{lagache03}, but consistent with those given by 
\citet{chary01} and \citet{takeuchi01a}.
\citet{perez_gonzalez05} adopted a linear relation between the SFR and the
dust luminosity given by \citet{kennicutt98} and obtained the evolution of 
the SFR density as $\rho_{\rm SFR}(\mbox{dust}) \propto (1+z)^{4.0\pm 0.2}$, 
which is in perfect agreement with the above result for the dust luminosity.
Therefore, this is a robust conclusion which does not depend on
the adopted LF functional shape for the fitting or the details of the 
conversion from dust emission to the SFR.

The fraction of obscured SFR density is also presented in 
Table~\ref{tab:lum_density}.
About half of the star formation activity is obscured in the local universe, 
while at $0.5 \la z \la 1.2$, about 70~\% of the total SFR is hidden by dust.
In particular, the obscured SFR fraction reaches more than $80~\%$ at 
$z\simeq 1$.
The result is consistent with a previous suggestion from the optical-to-FIR 
luminosity density ratio \cite[e.g.,][]{takeuchi01c}.
This result should substantially change the way we see the SFR history at
high-$z$.

\begin{acknowledgements}
We thank two anonymous referees for valuable suggestions which much improved
the paper.
We also deeply thank Emeric Le Floc'h for kindly providing their latest 
{\sl Spitzer} result before publication.
We are grateful to Takako T.\ Ishii and Hiroyuki Hirashita for helpful 
discussions.
Gemma Attrill and Alexie Leauthaud are sincerely thanked for their 
kind help for the improvement of English expressions.
TTT has been supported by the JSPS.
\end{acknowledgements}

\end{document}